\documentclass[a4paper,superscriptaddress,aps,prb,twocolumn,floatfix,citeautoscript,showkeys]{revtex4}

\usepackage{graphicx}
\usepackage{color}

\usepackage{amsmath}
\usepackage{amssymb}
\usepackage[english]{babel}
\usepackage[utf8]{inputenc}

\usepackage{bm}
\usepackage{url}
	
   \usepackage{amsfonts}
   \usepackage{multirow}
   \usepackage{nicefrac}
   \usepackage{graphicx,sidecap,wrapfig,lipsum,booktabs}
   \usepackage{epstopdf}

   \usepackage{amsmath}
   \usepackage{amssymb}	
   \usepackage{nicefrac}
   \usepackage{textcomp}
   \usepackage{verbatim}	

   \usepackage{tabularx}
   \usepackage{paralist}

   \usepackage{color}
	\usepackage[english]{babel}
\usepackage[colorlinks=true,linkcolor=blue,citecolor=blue,urlcolor=blue]{hyperref}

\usepackage{physics}
\graphicspath{{figures/}}
\newcommand{\GF}{Green's function}
\newcommand{\QP} {quasi-particle}

\newcommand{\CI}{Coulomb interaction}
\newcommand{\ham}{hamiltonian}
\newcommand{\HEG}{homogeneous electron gas}
\newcommand{\gowo}{G$_0$W$_0$}

\newcommand{\sgn}{\text{sgn}}

\newcommand{\imi}{\mathrm{i}}

\newcommand{\w}{\omega}
\newcommand{\beq}{\begin{equation}}
\newcommand{\eeq}{\end{equation}}
\newcommand{\bfk}{\mathbf{k}}

\begin{document}

\title{Cumulant Green's function calculations of plasmon satellites in bulk sodium: influence of screening and the crystal environment}

\newcommand{\lsi}{Laboratoire des Solides Irradi\'es, \'Ecole Polytechnique, CNRS, CEA,  Universit\'e Paris-Saclay, F-91128 Palaiseau, France}
\newcommand{\etsf}{European Theoretical Spectroscopy Facility (ETSF)}
\newcommand{\soleil}{Synchrotron SOLEIL, L'Orme des Merisiers, Saint-Aubin, BP 48, F-91192 Gif-sur-Yvette, France}
\newcommand{\seattle}{Department of Physics, University of Washington, Seattle, Washington 98195-1560, USA}

\author{Jianqiang Sky Zhou}
\email[]{jianqiang.zhou@polytechnique.edu}
\affiliation{\lsi}
\affiliation{\etsf}
\author{Matteo Gatti}
\affiliation{\lsi}
\affiliation{\etsf}
\affiliation{\soleil}
\author{J. J.  Kas}
\affiliation{\seattle}
\affiliation{\etsf}
\author{J. J. Rehr}
\affiliation{\seattle}
\affiliation{\etsf}
\author{Lucia Reining}
\affiliation{\lsi}
\affiliation{\etsf}
%
\begin{abstract}

We present \textit{ab initio} calculations of the photoemission spectra of bulk sodium using different flavors of the cumulant expansion  approximation for the Green's function. In particular, we study the dispersion and intensity of the plasmon satellites. We show that the satellite spectrum is much more sensitive to many details than the \QP\ spectrum, which suggests that the experimental investigation of satellites could yield additional information beyond the usual studies of the band structure. In particular, a comparison to the \HEG\ shows that the satellites are influenced by the crystal environment, although the crystal potential in sodium is  weak. Moreover, the temperature dependence of the lattice constant is reflected in the position of the satellites. Details of the screening also play an important role; in particular, the contribution of transitions from $2s$ and $2p$ semi-core levels influence the satellites, but not the \QP.  Moreover, inclusion of contributions to the screening beyond the RPA has an effect on the satellites. Finally, we elucidate the importance of the coupling of electrons and holes by comparing the results of the time-ordered (TOC) and the retarded (RC) cumulant expansion approximations. Again, we find small but noticeable differences. Since all the small effects add up, our most advanced calculation yields a satellite position which is improved with respect to previous calculations by almost one eV. This stresses the fact that the calculation of satellites is much more delicate than the calculation of a \QP\ band structure.

\end{abstract}
\date{\today}
\maketitle

\section{Introduction} 

Photoemission spectroscopy has become increasingly used to elucidate the electronic properties of materials, since it provides both \QP\ band structures, with information of one-particle-like excitations, and  satellite structures that reflect the coupling to bosonic excitations such as phonons, plasmons, magnons, etc.. \cite{Damascelli2013}
Accurate descriptions of photoemission spectra from \textit{ab initio} calculations have been a challenge for ages. 

Currently, the most widely used approach for moderately correlated materials is the GW approximation (GWA)   proposed by L. Hedin in 1965.\cite{Hedin-GW1965} 
In the GWA, the one-particle \GF\ is determined by a Dyson equation $G=G_H+G_H\Sigma_{xc} G$, 
where $G_H$ is the Hartree \GF, and $\Sigma_{xc}$ is a complex, non-local, and frequency dependent self-energy that is approximated as a convolution of the one-particle \GF\ $G$ and the dynamically screened \CI\ $W$, leading to $\Sigma_{xc}=\imi GW$. 
The GWA has become the state-of-the-art approach to compute  \QP\ band structures. However, one of its notable shortcomings is the poor description of the satellite structures in photoemission spectra.\cite{Martin2016}    
Since plasmons are the dominant structures in the inverse dielectric function $\epsilon^{-1}$ and hence in $W=\epsilon^{-1}v_c$, where $v_c$ is the bare \CI, one might suppose that plasmon satellites should be well described by the GWA.
However, this is not the case in practice. An example is the spurious prediction of a sharp plasmaron satellite, which has been contradicted experimentally.\cite{Lundqvist1967,Bergersen1972, Guzzo-prl2011,cumulant-Steven-2013} More in general, the GWA satellites due to plasmons  are generally too far from the \QP\ energy compared to the experiment.\cite{Aryasetiawan-cumulant1996,Aryasetiawan1998,cumulant-Ferdi-Al-experiment1999,Kheifets2003,Guzzo-prl2011,cumulant-Steven-2013,Guzzo2014}   

Alternatively, the cumulant expansion approximation (CEA) has been quite promising for giving a better description of plasmon satellites in photoemission spectra in a number of systems.\cite{Aryasetiawan-cumulant1996,Aryasetiawan1998,cumulant-Ferdi-Al-experiment1999,Aryasetiawan2000,Kheifets2003,Guzzo-prl2011,cumulant-Steven-2013,Gatti2013,Guzzo2014,Fabio-cumulant-2015,zhou-jcp2015,Lischner2015,Gumhalter2016,Nakamura2016,Verdi2017}
The CEA was inspired by the exact \GF\ of a electron-boson model {\ham} for a core level \cite{Langreth-quasiboson1970} and has been hence extensively used for core-level photoemission (see e.g. \cite{Chang1972,Chang1973,Huefner2003,DeGroot2008}), and also in other contexts, as for the electron-phonon interaction and the polaron problem (see e.g. \cite{Mahan1966,Dunn1975,Skinner1982,Hsu1984,Mahanbook}), or for modeling ultrafast electron dynamics (see e.g. \cite{Gumhalter2005,Gumhalter2006,Pavlyukh2016}).

The CEA  is represented by an exponential expression\cite{Kubo} of the \GF\ in the time-domain $G(t)=G_H(t)e^{C(t)}$, the expansion of which yields a Poisson series of  satellites in the spectral function $A(\omega)= \pi^{-1} |\Im G(\omega)|$,
consistent with experimental observations. Moreover, to lowest order in the screened interaction the cumulant function $C(t)$ can be expressed in terms of the GW self-energy, 
and it is therefore computationally no more demanding than the GWA itself.

The number of \textit{ab initio} CEA calculations to date is still relatively limited. Therefore, many details remain to be understood and settled.  
First, better agreement of CEA results with experiment is expected in insulators, semiconductors, or core levels of metals than in metal valence bands.\cite{zhou-jcp2015}  The reason is that the traditional time-ordered cumulant (TOC) is exact only in the limit of an approximated core-level \ham,\cite{Langreth-quasiboson1970} or for an approximation that decouples different orbitals.\cite{Guzzo-prl2011,zhou-jcp2015}  Both of these approximations assume that at zero temperature the occupation numbers are either $0$ or $1$, which is certainly not true close to the Fermi level of metals. 
A number of efforts have been made in order to go beyond the TOC    to describe  systems with partially occupied states. For example,   the retarded cumulant (RC)  approximation  was recently proposed,\cite{Josh-RC2014,Mayers2016} where both the \GF\ and the self-energy appearing in the CEA are replaced by their retarded counterparts. Consequently, while within the TOC unoccupied states do not produce satellites below the Fermi level, these additional features, which are a signature of coupling between occupied and unoccupied states, have been obtained in the {\HEG} by using the RC.\cite{Josh-RC2014}
Second, as pointed out above, the calculations rely on a GWA self-energy, which has been extensively studied for  calculations of \QP\ band structures. However, the insight gained from these studies is not necessarily transferable to the satellites, which are considerably enhanced by the CEA with respect to the GWA calculation. Indeed, our work shows that several effects influence the satellites, whereas they can be often overlooked for the \QP s. These include mild changes in the crystal environment and the lattice constant, the contribution of core levels, and the approximation used for the screening. 

We illustrate these points by performing both TOC and RC calculations for the valence photoemission spectrum of bulk sodium. Our most detailed calculation, which take into account all the aspects mentioned above, leads to an improvement of the satellite position of almost one eV with respect to previous calculations\cite{zhou-jcp2015}, as compared to experiment\cite{Hoechst1978}.

This paper starts in Sec. \ref{sec:theo} with a brief introduction to the theoretical framework, where those aspects are highlighted that are important for the subsequent analysis. In Sec. \ref{sec:TOC-RC} the results of the time-ordered and the retarded CEA are compared. Sec. \ref{sec:environment} discusses the effects of the crystal environment and the semi-core transitions on the spectra. Sec. \ref{sec:screening}  deals with the approximations used to calculate the screening. In Sec. \ref{sec:experiment} we compare our result to experiment. Finally, Sec. \ref{sec:conclusions} contains the conclusions. Computational details are relegated to  an appendix.

 \section{Theoretical framework} 
\label{sec:theo}

In this section we summarize the main theoretical ingredients needed for later analysis. 
In practical calculations, the cumulant expansion approximation for the \GF\ is combined with the GWA for the self-energy into the GW+C method\cite{Aryasetiawan-cumulant1996}. The traditional time-ordered version TOC 
for the diagonal matrix element of the Green's function $G$ in an occupied (hole) state  reads:
\beq
 G^{toc}(\tau) = \mathrm{i}\theta(-\tau) e^{-\mathrm{i}\varepsilon\tau}e^{C(\tau)} \, . 
\label{eq-toc-GF}
\eeq
Here the band $n$ and $\bfk$-point indexes have been dropped for simplicity (the TOC for an unoccupied state can be introduced in analogous way \cite{Aryasetiawan2000,Gatti2013}).
The \QP\ energy $\varepsilon$ is defined as: $\varepsilon = \varepsilon_H + \Re\Sigma_{xc}(\varepsilon)$,
with $\Sigma_{xc}$ the exchange-correlation self-energy calculated in an energy-self-consistent GW scheme. 
The TOC cumulant function $C(\tau)$ is obtained using the GW $\Sigma_{xc}$ as input: 
\beq
 C(\tau) = \frac{1}{\pi} \int_{-\infty}^{\mu-\varepsilon} \dd\omega \abs{\Im\Sigma_{xc}(\omega+\varepsilon)} \frac{e^{-\mathrm{i}\omega \tau} - 1}{\omega^2} \, . \label{eq-toc-C}
\eeq
Note that for hole states $\varepsilon$ is smaller than the Fermi energy $\mu$.
In a perturbative \gowo\ scheme for $\Sigma_{xc}$,  $\varepsilon$ would be a Kohn-Sham energy. The
\gowo\ approximation is however generally problematic for satellites \cite{Gatti2015,Bruneval2014}.
 
The retarded cumulant expansion can be obtained by simply replacing the time-ordered quantities (i.e., $G$ and $\Sigma_{xc}$) in Eqs. \eqref{eq-toc-GF} and \eqref{eq-toc-C} by their retarded counterparts:\cite{Josh-RC2014} 
\beq
 G^{rc}(\tau) = -i e^{-\mathrm{i}\varepsilon\tau}e^{C^R(\tau)} \, , \label{eq-rc-GF}
 \eeq
where 
 \beq
 C^R(\tau) = \frac{1}{\pi} \int_{-\infty}^{\infty} \dd\omega \abs{\Im\Sigma_{xc}^R(\omega+\varepsilon)} \frac{ e^{-\mathrm{i}\omega \tau}  - 1}{\omega^2} \, . \label{eq-rc-C}
\eeq
Here $\tau>0$ since we are interested in the removal sector; Of course, when time-ordered or retarded Green's functions are used consistently, the final result for observables should be exactly the same. Indeed, the retarded self-energy $\Sigma^R_{xc}$ can be replaced by the time-ordered one in Eqs. (\ref{eq-toc-GF}) or (\ref{eq-rc-C}), because the imaginary parts of retarded and time-ordered GW self-energies have the same absolute values. However, a difference may appear due to approximations with different consequences; indeed, the
difference in the cumulant functions is the integration range. This is due to a decoupling of electron and hole sectors in the derivation of the TOC.\cite{hedin-physscript,Almbladh1983,Aryasetiawan-cumulant1996,Guzzo-prl2011,zhou-jcp2015}  
As a consequence, the TOC only integrates the hole (or particle) part  of the self-energy (i.e., corresponding to $\w<\mu$), whereas the RC integrates both hole (lesser) and electron (greater) parts. 
Indeed, as with GW, the RC contains the coupling between occupied and unoccupied states, and is correct to first order in $W$.

In order to illustrate the physical meaning of the different terms in the cumulant function \eqref{eq-toc-C}, 
we consider a simple electron-boson model time-ordered self-energy: \cite{Gunnarsson-prb1994}
\beq
 \Sigma^{md}(\omega)= \frac{g^2/2}{\omega -\varepsilon_1 + \omega_p - \mathrm{i}\eta}+\frac{g^2/2}{\omega -\varepsilon_2 + \omega_p +\mathrm{i}\eta} \,, \label{eq-sigma-md2}
\eeq
where $g$ denotes the electron-plasmon coupling constant, $\omega_p$ is a non-dispersing plasmon  
energy, $\varepsilon_1<\mu$ and $\varepsilon_2>\mu$ are the energies of two electronic orbitals representing hole and electron state, respectively, and $\eta\rightarrow 0^+$ is an infinitesimal positive number.  
The imaginary part $\Im\Sigma^{md} = (\pi g^2/2) [\delta(\omega - \varepsilon_1 + \omega_p)-\delta(\omega - \varepsilon_2 - \omega_p)]$ 

contains one $\delta$-peak at $\omega = \varepsilon_1-\omega_p$ with weight $g^2/2$ 

and another one at $\omega = \varepsilon_2+\omega_p$ with weight $g^2/2$.

Using this model self-energy for the cumulant function \eqref{eq-toc-C}, we have
\beq
 C(\tau) =  \frac{g^2}{2\omega_p^2} e^{\mathrm{i}\omega_p \tau}   -\frac{g^2}{2\omega_p^2} \, .
\eeq

The physical meaning of each term in $C(\tau)$ becomes clear  in Eq. \eqref{eq-toc-GF}:
the first term generates a series of plasmon satellites at energies $\omega_p$ away from the \QP\ energy $\varepsilon^{toc}$.
The second term gives the \QP\ renormalization factor $Z^{toc} = \exp{-g^2/(2\omega_p^2)}$, 
which measures the spectral weight corresponding to the \QP\  excitation, whereas ($1-Z^{toc}$) goes into the rest of the spectral function, including satellites.

When $\Sigma^{md}$ is instead used in the $C^R$ \eqref{eq-rc-C}, we have for the matrix element of $G$ in the hole state with energy  $\varepsilon_1$ :
\beq
 C^R(\tau) =  \frac{g^2}{2\omega_p^2} e^{\mathrm{i}\omega_p \tau} +\frac{g^2}{2\tilde{\omega}_p^2}e^{-\mathrm{i}\tilde{\omega}_p \tau} -\frac{g^2}{2\omega_p^2}-\frac{g^2}{2\tilde{\omega}_p^2} \, ,
\eeq
where $\tilde{\omega}_p = \omega_p + \Delta$ with $ \Delta=\varepsilon_{2}-\varepsilon_{1}$ is the \QP\ energy difference between the two orbitals.

Two more terms appear in $C^R$ with respect to the time-ordered $C$ due to the electron part of $\Sigma^{md}$. The first new term generates a series of plasmon satellites at energies equal $\tilde{\omega}_p$ away from the \QP\ energy $\varepsilon_1$. Due to the minus sign in the exponential, these satellites are placed on the high energy side of $\varepsilon_1$. Therefore, the RC has satellite on both sides of the \QP\ peak in the spectral function. The second new term modifies the \QP\ renormalization factor, such that $Z^{rc} = \exp{-g^2/(2\omega_p^2)-g^2/(2\tilde{\omega}_p^2)}$. 

Analogously, the RC spectral function of the electron state (i.e., orbital with \QP\ energy $\varepsilon_2$) also contains satellites with energy below the Fermi energy. As a consequence, in order to have the complete RC electron removal spectrum, one also needs to sum also the spectral functions of the partially occupied electron states.    

The GW self-energy of a real system can still be written in the form of electron-boson coupling, but more poles appear.\cite{Hedin1999}
For each state $\ell$, its diagonal matrix element contains the sum of all valence and conduction states $j$ coupled with many different bosonic excitations $s$, and the time-ordered self-energy reads
\beq
 \Sigma_{xc}^{\ell\ell}(\omega) = \sum_{j,s\neq0}\frac{|V_{{\ell}j}^s|^2}{\omega - \varepsilon_j + (\omega_s -\mathrm{i} \eta)\sgn (\mu-\varepsilon_j)} \, . \label{Eq:model-sigma-Hedin}
\eeq
Here $\varepsilon_{j}$ are the \QP\ energies,  $\omega_s=E(N,s)-E(N,0)$ are the neutral excitation energies that correspond to the energy differences between the $N$-particle excited state $s$ and the $N$-particle ground state, and $V_{\ell j}^s$ are the fluctuation potentials, 
which determine the strength of the electron-boson coupling.
In the following, we will disentangle the different contributions by examining separately the various ingredients entering Eq. \eqref{Eq:model-sigma-Hedin}, 
and hence the cumulant expansions for $G$ in Eqs. \eqref{eq-toc-C}-\eqref{eq-rc-C}.

In a solid, it is convenient to analyze the loss function $L(\vb{q},\omega)$, which can be directly measured by inelastic x-ray scattering (IXS) or electron-energy-loss  spectroscopies (EELS):\cite{Schulke2007}
\beq
 L(\vb{q},\omega) = -\Im\epsilon^{-1}_M (\vb{q},\omega) = \frac{\epsilon_2(\vb{q},\omega)}{\epsilon_1^2(\vb{q},\omega)+\epsilon_2^2(\vb{q},\omega)} \, ,
\label{Eq:lossfunction}
 \eeq
where $\epsilon_M= \epsilon_1 + \mathrm{i} \epsilon_2$ is the complex macroscopic dielectric function.
The peaks of the loss function, which generally depend on the wavevector $\vb{q}$ (i.e., the experimental momentum transfer), 
correspond to the neutral excitations $\w_s$ in \eqref{Eq:model-sigma-Hedin}. 
In particular, the plasmon energies $\omega_{pl}(\vb{q})$ 
correspond to the peaks in the loss function for which $\epsilon_1(\vb{q},\omega_{pl}(\vb{q}))=0$.

Eq. \eqref{Eq:model-sigma-Hedin}
shows that the self-energy is an average over the couplings of the single-particle states $\ell$ with the plasmons (and other electron-hole excitations)
at all momentum transfers $\vb{q}$.
As a consequence, the plasmon energy $\omega_{pl}(\vb{q}=0)$ is in general  different from the plasmon satellite energies $\omega_{ps}$ 
in the spectral function, which for each state are defined as the energy distance between the \QP\ and first plasmon satellite. 

In the GWA the inverse dielectric function $\epsilon^{-1}$ and the loss function are often calculated within the random-phase approximation (RPA).
However, one may go beyond the RPA by using time-dependent density-functional theory (TDDFT)\cite{Runge1984,ullrich-book}, where the solution of the Dyson equation for the polarizability $\chi=\chi_0+\chi_0(v_c+f_{xc})\chi$ yields $\epsilon^{-1}=1+v_c\chi$. While the RPA corresponds to setting the exchange-correlation kernel $f_{xc}$ to 0 and evaluating the independent-particle polarizability $\chi_0$ in some mean-field approximation, 
the most widely used approximation in TDDFT is the adiabatic local-density approximation (ALDA)\cite{Zangwill1980,Petersilka1996}.
In general, the ALDA yields plasmon spectra in better agreement with EELS and IXS experiments than the RPA.\cite{Onida2002,Botti2007}
Therefore we will investigate whether the ALDA also improves plasmon satellites in photoemission spectra.

\section{Time-ordered versus retarded cumulant approximation}
\label{sec:TOC-RC}
\begin{figure}
\begin{center}
\includegraphics*[width=0.9\columnwidth, angle=0]{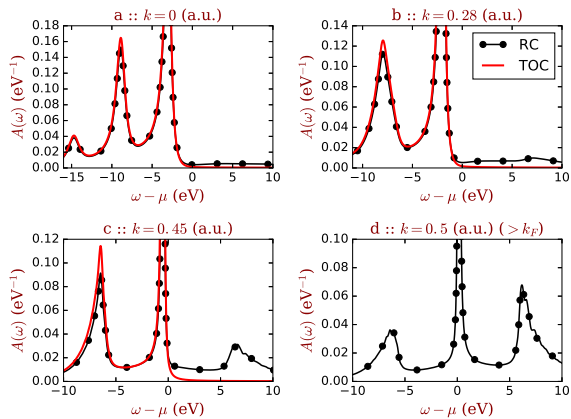}
\end{center}
\vspace{-15pt}
\caption{$\vb{k}$-resolved spectral functions $A(\vb{k},\omega)$ for the sodium valence band along  $\Gamma$N using TOC (red solid curve) and RC (black dotted curve). The Fermi wavevector is $k_{F}= 0.49$ a.u..}
\label{Fig:spf-na-toc-rc}
\end{figure}

\begin{figure}
\begin{center}
\includegraphics*[width=0.9\columnwidth, angle=0]{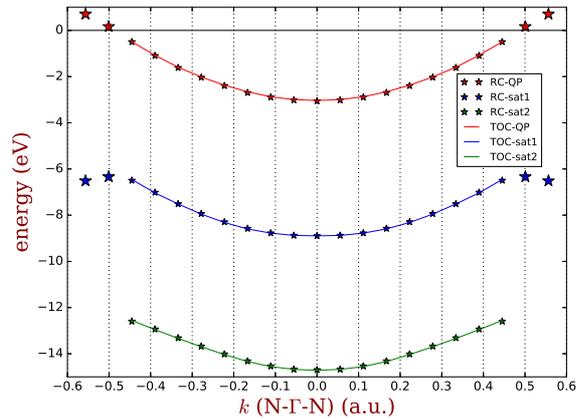}
\end{center}
\vspace{-15pt}
\caption{The dispersion of \QP s (red), first (blue) and second (green) satellites in TOC (solid curves) and RC (stars). The stars with double size are satellites from states above the Fermi level $\mu$.}
\label{Fig:band-na-toc-rc}
\end{figure}

The RC has been applied to the homogeneous electron gas, \cite{Josh-RC2014} but it remains interesting to investigate whether the RC improves over the TOC 
for the spectral function of a real metal like sodium.
To this end, we perform \textit{ab initio} TOC and RC calculations for bulk sodium, using the computational ingredients summarized in App. \ref{sec:compdet}. The TOC and RC results are compared in Fig. \ref{Fig:spf-na-toc-rc},  
which shows the $\vb{k}$-resolved spectral functions $A(\vb{k},\w)$ along the $\Gamma$N direction for the sodium valence band,  
crossing the Fermi level at $k_F \sim 0.49$ a.u..
At $\vb{k}=\Gamma$, which is at the bottom of band, and for states close to it, 
the TOC and the RC spectral functions   are very similar for $\omega < \mu$.
In agreement with previous TOC calculations\cite{Aryasetiawan-cumulant1996,zhou-jcp2015},
there is a prominent \QP\ peak which has a parabolic dispersion (see Fig. \ref{Fig:band-na-toc-rc})
and two satellites that follow the \QP\ dispersion at a distance of $\sim 5.84$ eV  and $\sim 5.8$ eV, respectively. This similar dispersion is analogous to the situation in silicon, which has been investigated in \cite{Fabio-cumulant-2015,Lischner2015}. 
The satellites are slightly more intense in the TOC than in the RC, as the renormalization factors are different in the two cases.
For $\omega > \mu$, the RC  displays a non-zero spectral weight, while  the TOC is always 0 by definition.
This tail in the RC comes from the integration of the electron part of $\Im\Sigma_{xc}$, which is present in the RC but not in the TOC.
By approaching $k_F$ the differences between TOC and RC spectral functions become significant: 
the unoccupied part of the RC spectral function becomes larger, also showing a pronounced satellite at about $6$ eV above $\mu$. 
For $k \sim k_F$ the RC is symmetric around $\mu$.
Finally for $k>k_F$ [see Fig. \ref{Fig:spf-na-toc-rc}(d)] we show only the RC, since this matrix element of the hole TOC is zero.
Interestingly, for $k>k_F$  the RC still has a satellite for $\w<\mu$, 
which might be measurable by photoemission experiments.

\begin{figure}
\begin{center}
\includegraphics*[width=0.9\columnwidth, angle=0]{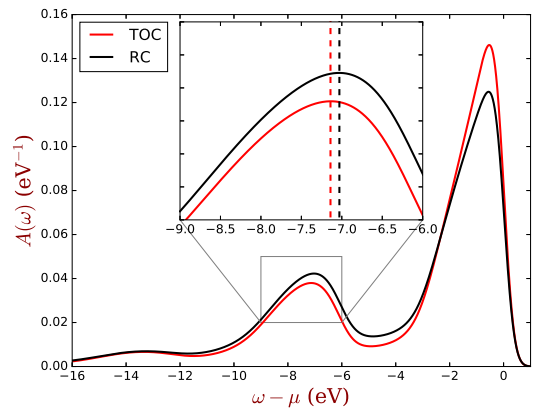}
\end{center}
\vspace{-15pt}
\caption{(Main) The total $\vb{k}$-summed spectra for the sodium valence band in both TOC (in red) and RC (in black), multiplied by the room temperature Fermi function. (Inset) Zoom on the first satellite. The two dashed vertical lines mark the positions of the maximum of each satellite: their distance is $0.11$ eV. 
}
\label{Fig:spftot-na-toc-rc}
\end{figure}

Fig. \ref{Fig:band-na-toc-rc} shows the dispersion of the band and the satellites. 
While for occupied states the QP and the satellites have the same parabolic dispersion, for unoccupied  states we find that the satellites in the RC spectral function below the Fermi level, 
which are denoted by stars in the figure,
do not follow the parabolic dispersion of the \QP\ band, becoming more flat and 
even inverting the curvature. This behavior can be understood using the model equations (\ref{eq-rc-GF}), (\ref{eq-rc-C})  and (\ref{eq-sigma-md2}), by varying the energy $\varepsilon_2$ in order to simulate the dispersion of the empty state. 

Fig. \ref{Fig:spftot-na-toc-rc} shows the valence spectral functions summed over the first two bands, integrated over all $\vb{k}$ in the Brillouin zone, and multiplied by the Fermi function for T=300 K together with a 0.3 eV Gaussian broadening.
While qualitatively similar, 
the TOC and RC display small quantitative differences for both the QP peak at the Fermi level and the satellites (see the zoom around the first satellite in the inset of Fig. \ref{Fig:spftot-na-toc-rc}).
Notably the maximum of the first satellite is more intense and closer to the QP peak in the RC compared to the TOC. 
The differences between TOC and RC are due to the different renormalization factors and to the satellites  of the unoccupied states for $\w<\mu$, which are present only in the RC spectral functions.

The maximum of the RC satellite has a binding energy that is 0.11 eV smaller than that of the TOC, bringing it into better agreement with experiment. 
We conclude that the RC leads to some small, but visible changes in the valence photoemission spectra of a metal such as sodium. Since the RC contains additional physics, one may expect that this approximation is better than  the TOC.
In the following we will present only RC spectral functions.

\section{Environment effects on the plasmon satellites}
\label{sec:environment}

In this section we will study various contributions that have a small, but visible influence on the satellites, while they do not affect the \QP s.  It should be noted that all effects discussed here lead to changes of the same sign, such that they add up and finally have a non-negligible impact on the spectra.

\subsection{The lattice potential: comparing sodium and the homogeneous electron gas}
\label{Sec:Na-HEG}

Sodium is the closest realization of the homogeneous electron gas (HEG) model: 
the potential due to the ionic lattice introduces only a very small perturbation of the ideal HEG,
the valence-band dispersion remains close to parabolic and the Fermi surface close to spherical.
The spectral function of the HEG has been previously calculated using both the TOC\cite{Aryasetiawan-scHEG1997,Caruso-HEG2016,Johannes-2DHEG2016} and the RC\cite{Josh-RC2014,Mayers2016} that we employ here.
By comparing sodium and the HEG with the same electron density, here
we can additionally establish whether the lattice potential influences the \QP\ and satellite properties in the same way.

\begin{figure}
\begin{center}
\includegraphics*[width=0.9\columnwidth, angle=0]{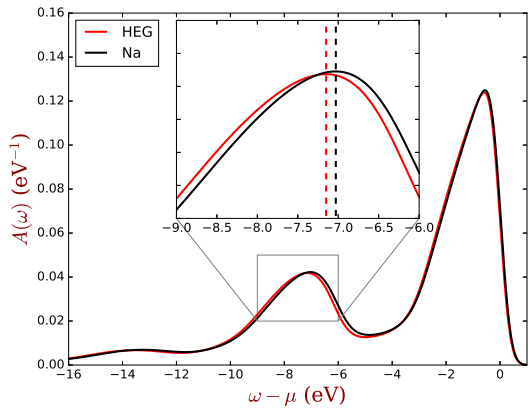}
\end{center}
\vspace{-15pt}
\caption{(Main) The total $\vb{k}$-summed spectra of HEG (in red) and sodium (in black), multiplied by the Fermi function. (Inset) Zoom on the first satellite.
The two dashed vertical lines mark the positions of the maximum of each satellite: their distance is $0.12$ eV. 
}
\label{Fig:spftot-na-heg}
\end{figure}

\begin{figure}
\begin{center}
\includegraphics*[width=0.9\columnwidth, angle=0]{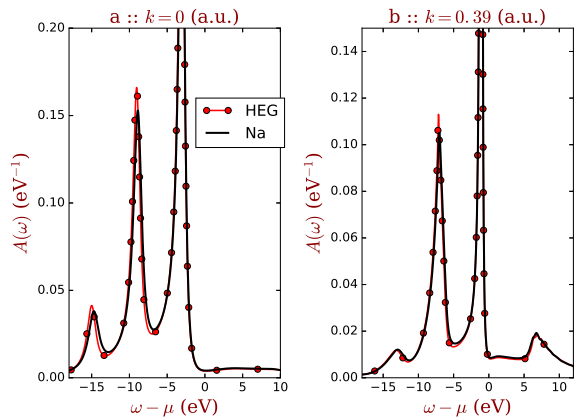}
\end{center}
\vspace{-15pt}
\caption{$\vb{k}$-resolved spectral functions A$(\vb{k},\omega)$ of sodium (red curve with circles) and the HEG (black curve) at the $\Gamma$ point and close to Fermi level ($k_{F}\sim 0.49$ a.u.).}
\label{Fig:spf-na-heg}
\end{figure}

\begin{figure}
\begin{center}
\includegraphics*[width=0.9\columnwidth, angle=0]{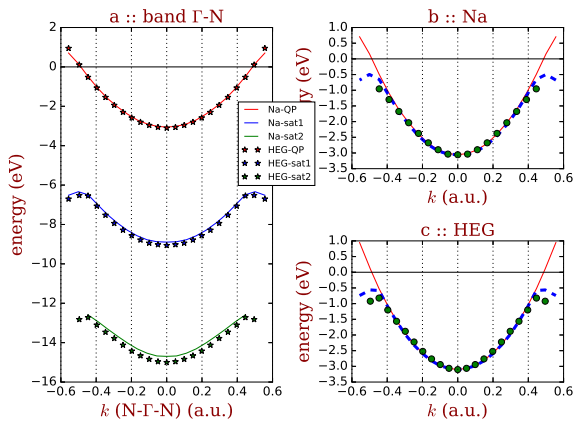}
\end{center}
\vspace{-15pt}
\caption{(a): Dispersion of \QP s (in red), first (in blue) and second (in green) satellite of sodium (solid curves) and of the HEG (stars) along the N$\Gamma$N direction.  (b)-(c) Comparison of the \QP\ (red solid curves), first satellite (blue dashed curve), and second satellite (green dots) dispersions of (b) sodium and (c) the HEG. The satellite energies have been shifted in order to align all energies at the $\Gamma$ point.}
\label{Fig:band-na-heg}
\end{figure}

The integrated spectral functions for Na and the HEG, which are displayed in Fig. \ref{Fig:spftot-na-heg},
are very similar for the \QP\ peak at the chemical potential $\mu$ whereas their satellites are slightly  different (see the zoom in the inset of Fig. \ref{Fig:spftot-na-heg}).
In the HEG the satellite has a larger distance from the  \QP\ than in Na, resulting in a larger binding energy.
This is confirmed by comparing in Fig. \ref{Fig:spf-na-heg} the dispersion of the $\vb{k}$-resolved spectra along 
$\Gamma$N. 
While the \QP\ bands overlap entirely in the two cases, 
at each $\vb{k}$ point the distance between the \QP\ and the first satellite is larger in the HEG than in Na. This difference is almost twice as big for the second satellite [see Fig. \ref{Fig:band-na-heg}(a)]. We also note that the largest differences occur around the $\Gamma$ point at the bottom at the band, while around the Fermi level the satellite positions get closer.

For a better comparison, Fig. \ref{Fig:band-na-heg}(b)-(c) shows for both  sodium and the HEG the dispersion of the \QP\ band and the plasmon satellites, where the satellite energies have been shifted such that they coincide with the 
\QP\ at $\Gamma$. As already found in sodium (see Fig. \ref{Fig:band-na-toc-rc}), also in the HEG  at the bottom of the band the satellite band follows the parabolic dispersion of the {\QP}. When the state is instead close to Fermi level, there is an abrupt change, yielding a flat dispersion and a downwards bending. Since this feature is in common for the HEG and Na, this property of the satellite dispersion must be due to the electronic interaction, 
while the differences in the satellite energies between the HEG and Na are caused by the lattice potential.

\begin{figure}
\begin{center}
\includegraphics*[width=0.9\columnwidth, angle=0]{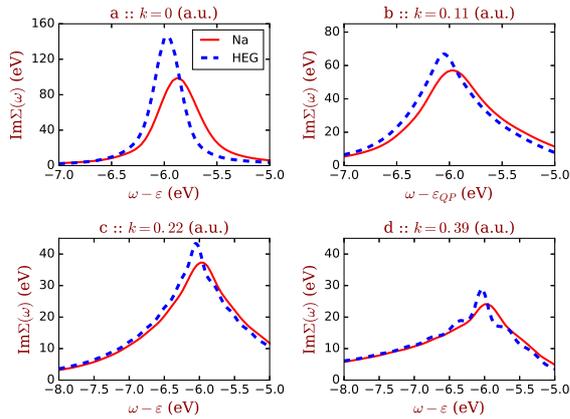}
\end{center}
\vspace{-15pt}
\caption{The shifted imaginary part of self-energy of sodium (red solid lines) and HEG (blue dashed lines) at different $\bfk$-points along $\Gamma$N in the sodium Brillouin zone. Only the removal part $\w<\mu$ is shown.  The Fermi wavevector is at $k_F= 0.49$ a.u.. }
\label{Fig:beta-na-heg}
\end{figure}

In order to understand the origin of these differences, let us analyze  the imaginary part of the self-energy $\Im\Sigma_{xc}$ that enters  Eq. \eqref{eq-rc-C}, shown in Fig.  \ref{Fig:beta-na-heg}.
In both sodium and the HEG, $\Im\Sigma_{xc}$ is characterized by a single peak, which in the HEG is located at larger distances from the corresponding \QP\ peak than in Na (note that the energy scale in the figure is given relative to the \QP\ energy). 
This explains why the satellites in the spectral functions are at higher binding energies in the HEG.

Approaching the Fermi level, the difference between sodium and the HEG decreases, while the peak becomes broader and asymmetric. 
The shape of $\Im\Sigma_{xc}$ can be directly linked, through  Eq. \eqref{Eq:model-sigma-Hedin}, 
to the  parabolic valence band dispersion in Na and in the HEG.
Typically, for a given bosonic excitations $s$, the dominant contribution to the sum over states $j$ is selected by the coupling matrix elements $V_{\ell j}^s$ and stems from states that are close, i.e. ${\rm Im}\,\Sigma_{\rm xc}^{\ell\ell}$ is dominated by contributions with $\abs{\vb{k}_j -\vb{k}_{\ell}} < \Delta$.
When $\vb{k}_{\ell}$ is at the bottom of the parabolic band, i.e. close to $\Gamma$ where the band is relative flat, 
neighboring states $\vb{k}_j$ for which $V_{\ell j}^s$ is significantly different from zero have energies $\varepsilon_j$ 
very close each other.
 
As the result, $\Im\Sigma_{xc}^{\ell\ell}$ for such a state $\ell$ has a sharp peak around $\varepsilon_{\ell}-\omega_s$.
Instead, when $\vb{k}_{\ell}$ is away from $\Gamma$, where the band has a steeper slope, 
 $\Im\Sigma_{xc}^{\ell\ell}$ is different from zero in  a wider energy range.
At the same time, it becomes more asymmetric, developing a long tail on the low-energy side. The reason for the asymmetry is the availability of energies: close to the Fermi level, there are fewer  occupied state $j$ with energy $\varepsilon_j > \varepsilon_{\ell}$, whereas many  states with smaller energies $\varepsilon_j<\varepsilon_{\ell}$
contribute to the low-energy tail of the peak. Going towards the bottom of the valence band, the spectral weight continuously moves towards the high-energy side of the peak. At the bottom of the band, however, the asymmetry is hidden by the fact that the peak is sharp.
Of course, this is a qualitative analysis, since the coupling with bosonic excitations of different character and energies $\omega_s$ that are summed up to form the self-energy complicates the picture.

\begin{figure}
\begin{center}
\includegraphics*[width=0.9\columnwidth, angle=0]{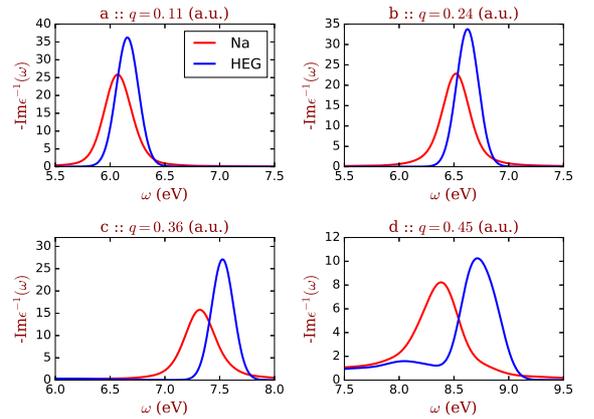}
\end{center}
\vspace{-15pt}
\caption{ The RPA loss functions of sodium (red curve) and HEG (blue curve) at different momentum transfers $q$ (in a.u.)} 
\label{Fig:loss-na-heg}
\end{figure}

Finally, in order to understand why the peak position of $\Im\Sigma_{xc}$ in the HEG is always further from the \QP\ than in Na, 
we compare the loss functions, 
which are shown in Fig. \ref{Fig:loss-na-heg} as a function of momentum transfer $q$.
For $q$ smaller than the wavevector $q_c \sim  0.45$ a.u, the peak in the loss function corresponds to a plasmon resonance, for which $\epsilon_1=0$ [see Eq. \eqref{Eq:lossfunction}].
Above $q_c$ the plasmon enters the electron-hole continuum where the loss function is dominated by $\epsilon_2$.
In agreement with Ref. \onlinecite{Cazzaniga-loss2011}, the HEG shows larger plasmon energies than sodium at all momentum transfers.
As $q$ increases, the difference becomes larger and larger: the plasmon in sodium is more and more affected by band-structure effects and short-range spatial inhomogeneities in the charge response become more apparent. These observations suggest that low-density regions have a stronger influence on the plasmon energy of an inhomogeneous material than high-density regions, such that the resulting plasmon energy is smaller than what one would expect from the average density.

This difference in the plasmon energies explains why the plasmon satellite has a larger binding energy in the HEG than in Na.
Since the difference in the peak position of $\Im\Sigma_{xc}$ between sodium and the HEG is always smaller than $0.2$ eV (see Fig. \ref{Fig:beta-na-heg}),
we can conclude that  the loss functions at small momentum transfers (i.e. $q \lesssim 0.3$ a.u., where the loss functions of the HEG and Na are similar), are those which contribute mostly to $\Im\Sigma_{xc}$ and hence to the position of the plasmon satellite in the spectral functions.

\subsection{Thermal expansion}
\label{Sec:5k-293k}

\begin{figure}
\begin{center}
\includegraphics*[width=0.9\columnwidth, angle=0]{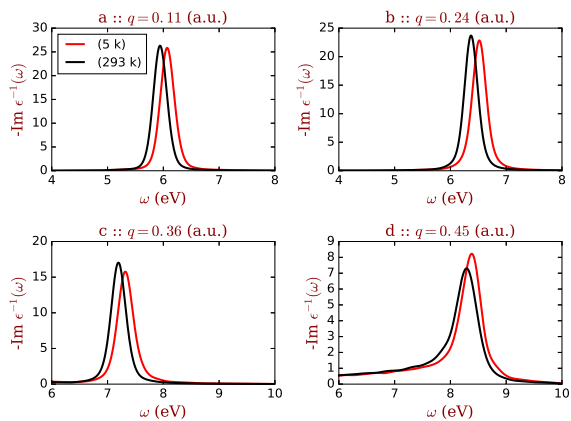}
\end{center}
\vspace{-15pt}
\caption{RPA Loss functions of sodium calculated with lattice parameters corresponding to 5 K (red curve) and 293 K (black curve) at different momentum transfers $q$ (in a.u.). }
\label{Fig:loss-5k-293k}
\end{figure}

\begin{figure}
\begin{center}
\includegraphics*[width=0.9\columnwidth, angle=0]{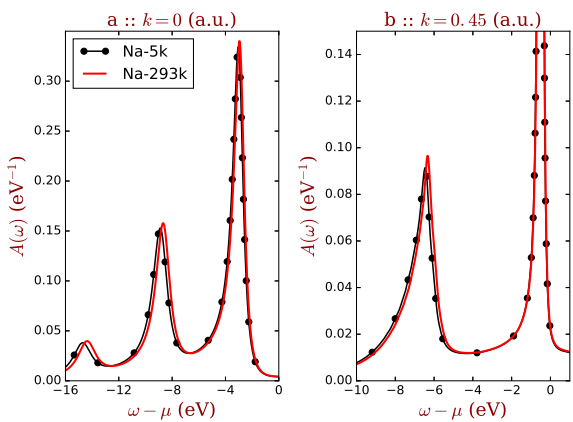}
\end{center}
\vspace{-15pt}
\caption{$\vb{k}$-resolved spectral functions of sodium  along the $\Gamma$N direction using lattice parameters corresponding to 5 K (black curve with dots) and 293 K (red solid curve) at the $\Gamma$ point and close to Fermi level ($k_F =$ 0.49 a.u.)} 
\label{Fig:spfk-5k-293k-1}
\end{figure}

\begin{figure}
\begin{center}
\includegraphics*[width=0.9\columnwidth, angle=0]{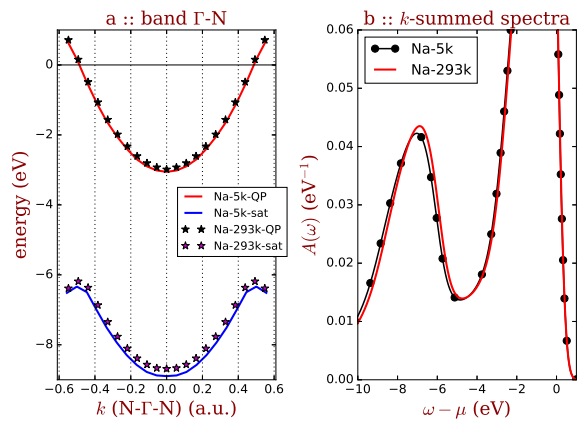}
\end{center}
\vspace{-15pt}
\caption{ (a) The QP and plasmon satellite dispersions along $\Gamma$N. The QP and first satellite energies of Na at 5K are represented in red and blue solid curves, respectively. For Na at 293K QP and satellite energies are stars. 
(d) The total spectra summed over $\vb{k}$-points and two bands, using lattice parameters corresponding to 5 K (black curve with dots) and 293 K (red solid curve). 
}
\label{Fig:spfk-5k-293k-2}
\end{figure}

\begin{figure}
\begin{center}
\includegraphics*[width=0.9\columnwidth, angle=0]{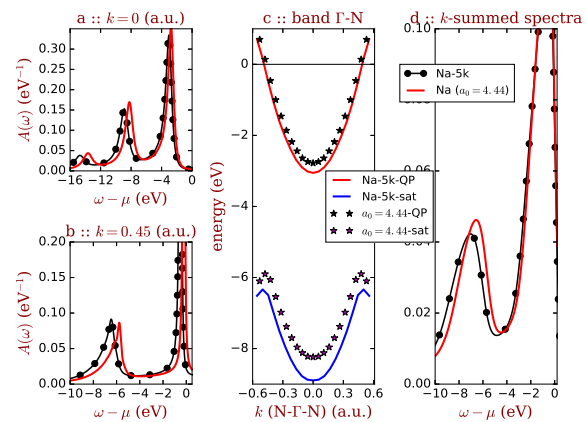}
\end{center}
\vspace{-15pt}
\caption{Comparison between Na with lattice parameter at 5 K ($a_0=4.23$ \AA) and Na with artificially expanded lattice parameter ($a_0=4.44$ \AA): $\vb{k}$-resolved spectral functions  (a) at the $\Gamma$ point and (b) close to the Fermi level; (c) Band and satellite dispersions; (d) $\vb{k}$-integrated spectral function.}
\label{Fig:spftot-5k-500k}
\end{figure}

The results above have been obtained with calculations performed at temperature $T=0$, and with a lattice parameter for sodium of $4.227$ {\AA}, which is the experimental result measured at $T=5 K$. However, experiments are often performed at room temperature, $T=293 K$. While we do not expect a major influence of the electronic temperature besides the Fermi function in the spectra, thermal expansion may play an important role. Indeed,  
by increasing the temperature from 5 K to room temperature, the lattice parameter of sodium  changes considerably, from $4.227$ {\AA} to $4.29$ \AA  \cite{Wyckoff-lattice1963}.
Since the plasmon energy at $\vb{q}=0$ is approximatively proportional to the square root of the electronic density, 
we expect that with the decrease of density at higher temperature, 
the plasmon energy decreases as a consequence of the lattice thermal expansion.
Indeed, in Fig. \ref{Fig:loss-5k-293k} we find that for all momentum transfers 
the plasmon resonance is located at lower energies in the loss function calculated with the room-temperature lattice parameter than in the 5 K result.

Extrapolating from the comparison between Na and the HEG in Sec. \ref{Sec:Na-HEG}, 
one should expect a similar effect on the spectral functions.
Indeed, for all $\vb{k}$ points, the plasmon satellites in Figs. \ref{Fig:spfk-5k-293k-1}-\ref{Fig:spfk-5k-293k-2} have smaller binding energies at room temperature than at 5 K. Again, the satellites 
are more affected by the thermal expansion than the QP peaks, 
which remain almost unchanged.

This trend is confirmed by a calculation where we have artificially expanded the lattice parameter to $4.44$ \AA\ for the sake of demonstration.
Fig. \ref{Fig:spftot-5k-500k} shows that the satellite band moves much closer to the QP band, which does change, but to a much lesser extent: while the QP bandwidth is reduced by 0.28 eV, the binding energy of the maximum of the satellite peak decreases by 0.46 eV.

\subsection{Core polarization \label{Sec:valence-core}}

Transition from shallow core levels to empty states are known to affect the loss function also at lower energies, 
i.e., in the energy range of valence transitions.\cite{Taut1987,Sturm1990,Quong1993}
Since we have found that the satellite in the spectral function is very sensitive to small changes of the plasmon properties, 
here we analyze whether those core polarization effects have an influence also on the valence plasmon satellites in the spectral function of sodium.

\begin{figure}
\begin{center}
\includegraphics*[width=0.9\columnwidth, angle=0]{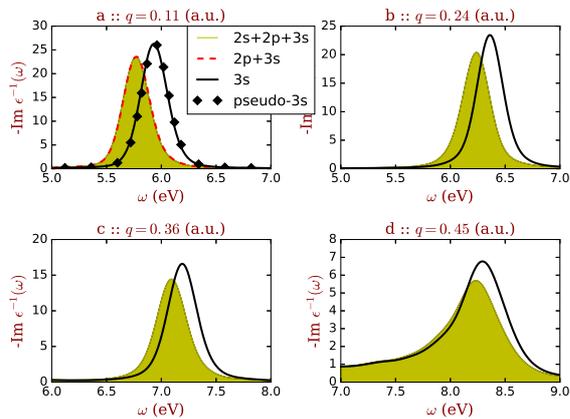}
\end{center}
\vspace{-15pt}
\caption{The loss functions $-\Im\epsilon^{-1}$ including different transitions. The yellow filled curve contains all transitions from $2s$, $2p$, $3s$ states. The red dashed and black solid curves contain transitions from $2p+3s$ and $3s$ only, respectively. The black diamonds are  calculated using a pseudopotential containing only $3s$ electrons as valence states. 
}
\label{Fig:loss-core-valence}
\end{figure}

In order to investigate how the $2s$ and $2p$ core states affect the loss function we have used two different pseudopotentials: 
one that has only $3s$ as valence electrons and another where also $2s$ and $2p$ are explicitly included in the calculations.\cite{Gatti2010}
First of all, we have to make sure that the errors inherent in the pseudopotential approach do not bias our conclusions. To this end, we have verified that the two pseudopotentials give the same result when only excitations from $3s$ states are taken into account. This is indeed the case, as one can see from the comparison of the two pseudopotential results (black diamonds and black curve) in Fig. \ref{Fig:loss-core-valence}(a).
In the next step, we add transitions from $2s$ and $2p$ core levels to the calculations. This leads to the yellow shaded curves, 
which are redshifted with respect to the black curves for all momentum transfers [see Fig. \ref{Fig:loss-core-valence}(a)-(d)].
This effect is mainly due to the $2p$ electrons: results with [yellow shaded curves] or without [red curve in Fig. \ref{Fig:loss-core-valence}(a)] the $2s$ are indistinguishable. 

\begin{figure}
\begin{center}
\includegraphics*[width=0.9\columnwidth, angle=0]{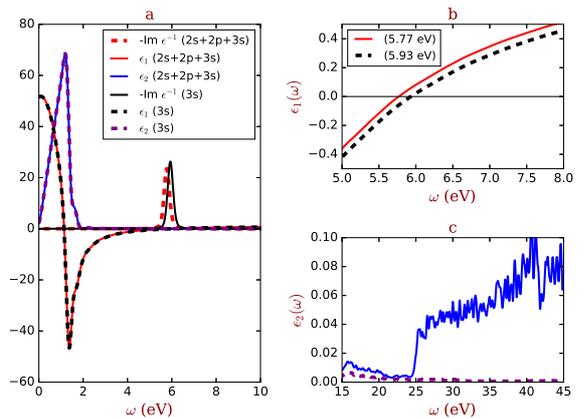}
\end{center}
\vspace{-15pt}
\caption{ (a) The loss functions $-\Im\epsilon^{-1}$ containing the core transitions (red dashed curve) and transitions of valence states only (black solid curve), together with their real ($\epsilon_1$) and imaginary ($\epsilon_2$) parts at $q = 0.11$ a.u.. (b) Zoom around the plasmon energy for $\epsilon_1$ (c) Zoom around the core-level contributions for $\epsilon_2$. 
}
\label{Fig:abs-core-valence}
\end{figure}

\begin{figure}
\begin{center}
\includegraphics*[width=0.9\columnwidth, angle=0]{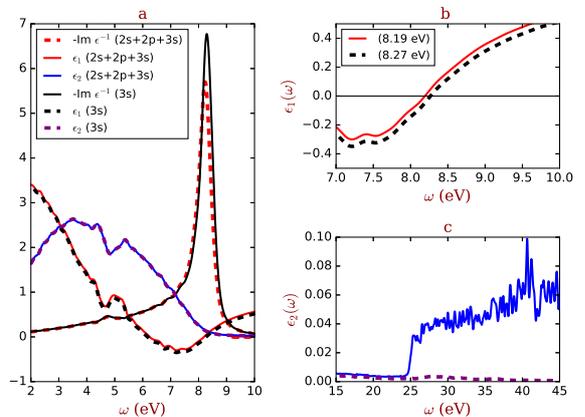}
\end{center}
\vspace{-15pt}
\caption{
Same as Fig. \ref{Fig:abs-core-valence}, but for $q = 0.45$ a.u..}
\label{Fig:abs-core-valence-2}
\end{figure}

To understand the origin of the redshift of the loss function, the real and imaginary parts of the dielectric functions at momentum transfers $q=0.11$ a.u. and $q=0.45$ a.u. are shown in Figs. \ref{Fig:abs-core-valence} and \ref{Fig:abs-core-valence-2}, respectively.
When the transitions from core levels are included in the calculation, at smaller energies $\epsilon_2$ is unchanged, but at energies larger than 25 eV, which corresponds to the core-level binding energies, a new structure appears.
As a consequence, through the Krames-Kronig relation, $\epsilon_1$ is affected on a wider energy range. In particular the position of its crossing with the zero axis is shifted, which changes the plasmon peak in the loss function.
This effect is smaller at larger momentum transfers.

\begin{figure}
\begin{center}
\includegraphics*[width=0.9\columnwidth, angle=0]{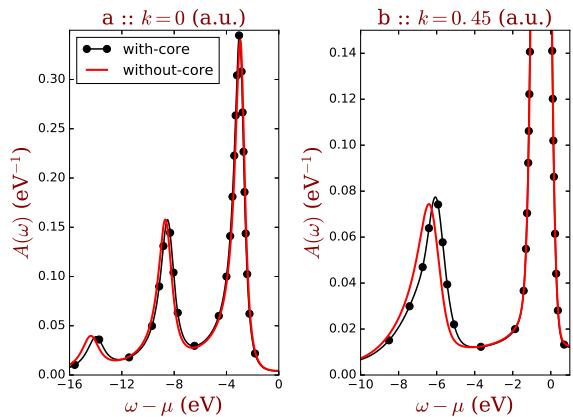}
\end{center}
\vspace{-15pt}
\caption{$\vb{k}$-resolved spectral functions A$(\vb{k},\omega)$ of sodium taking into account the core polarization (black curve with dots) and without core polarization (red curve) (a) at the $\Gamma$ point and (b) close to Fermi level ($k_{F}= 0.49$ a.u.).} 
\label{Fig:spf-core-valence}
\end{figure}

\begin{figure}
\begin{center}
\includegraphics*[width=0.9\columnwidth, angle=0]{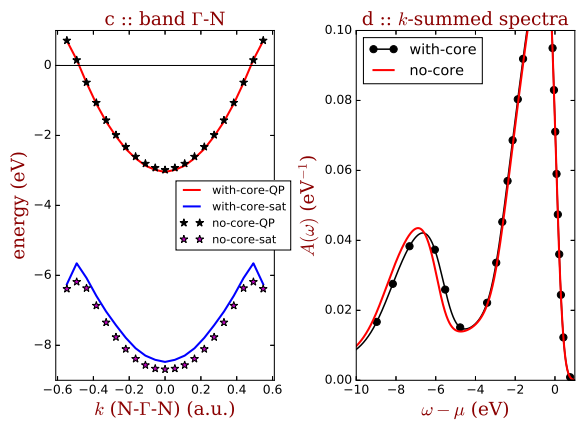}
\end{center}
\vspace{-15pt}
\caption{(a) Band and satellite dispersions along $\Gamma$N and (b) $\vb{k}$-integrated spectral function, calculated with or without the core polarization contribution.} 
\label{Fig:band-core-valence}
\end{figure}

The core-polarization effect in the loss functions influences the spectral functions for the sodium valence band (see Fig. \ref{Fig:spf-core-valence}).
As in the previous cases, the QP peak is affected in a negligible way, while the plasmon satellite energy in the $\vb{k}$-integrated spectral function (see Fig. \ref{Fig:band-core-valence}) is reduced by about 0.23 eV by including the core-level transitions in the screening calculation.

Altogether, the results presented in the this and previous two sections 
clearly illustrate that the plasmon satellite 
is very sensitive to all the changes of the environment 
surrounding the \QP\ excitation. 
The lattice potential, 
the change in the lattice parameter,
and the polarization from the core electrons, have a much stronger 
influence on the plasmon satellites than on the \QP\ peaks.
This finding is consistent with the observation made for the comparison of graphene and graphite in Ref. \onlinecite{Guzzo2014}: also in that case it was found that the presence of neighboring graphene planes in graphite affects more the satellite than the QP spectra.
This implies that plasmon satellites in photoemission spectra
are powerful ``detectors'' for small variations of a material, and that measuring and analyzing the satellites in photoemission spectra, in addition to the \QP\ peaks, may give additional precious information.

\section{The screened interaction beyond the RPA \label{Sec:rc}}
\label{sec:screening}

\begin{figure}
\begin{center}
\includegraphics*[width=0.9\columnwidth, angle=0]{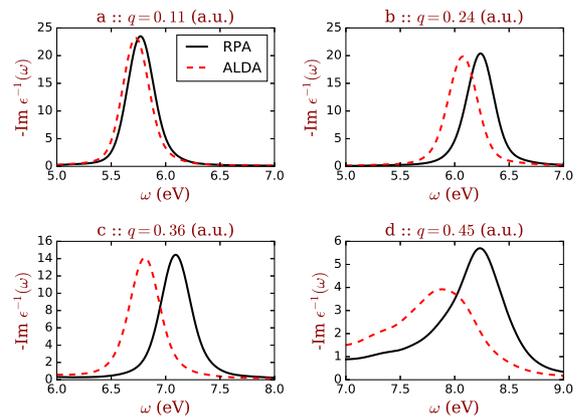}
\end{center}
\vspace{-15pt}
\caption{The loss functions $-\Im\epsilon^{-1}$ of Na calculated in RPA (black solid lines) and ALDA (red dashed lines) at different momentum transfers $\vb{q}$ in a.u..
} 
\label{Fig:loss-RPA-TDLDA}
\end{figure}

\begin{figure}
\begin{center}
\includegraphics*[width=0.9\columnwidth, angle=0]{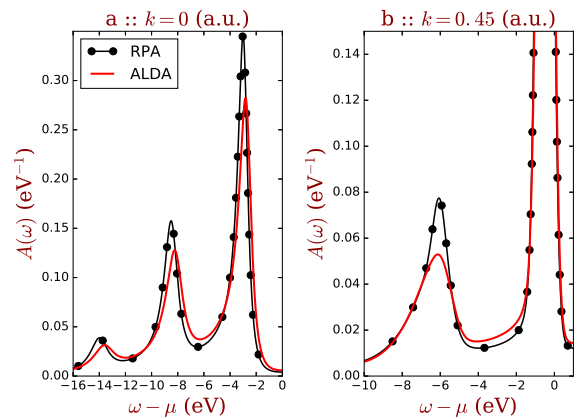}
\end{center}
\vspace{-15pt}
\caption{$\vb{k}$-resolved spectral functions $A(\vb{k},\omega)$ of sodium using RPA  (black dotted curve) and ALDA screening (red curve) (a) at the $\Gamma$ point and (b) close to Fermi level ($k_{F}= 0.49$ a.u.).} 
\label{Fig:spf-RPA-TDLDA}
\end{figure}

In the GWA the screening given by the inverse dielectric function $\epsilon^{-1}$ is usually calculated at the level of the RPA.
However, previous studies have shown that in sodium, like in other materials, 
the ALDA yields loss functions in better agreement with IXS experiments, \cite{Cazzaniga-loss2011,Quong1993}
since it leads to a redshift of the plasmon energy that increases with the momentum transfer.
This is confirmed by the results reported in Fig. \ref{Fig:loss-RPA-TDLDA}.
One would therefore expect that the choice of the ALDA or the RPA for the calculation of the screening should significantly affect the plasmon satellites.
On the other hand, our previous analysis shows that the satellite position in the spectral function is mainly determined by the plasmon energy at small momentum transfers, where the difference between the RPA and the ALDA, and the difference between the RPA and experiment, are minor. This rises the question of how important it is to go beyond the RPA in the calculation of plasmon satellites, and whether the calculation that yields loss functions in better agreement with IXS measurements also yields plasmon satellite spectra in better agreement with photoemission experiments. This is a non-trivial question, and we can only give evidence, since the quality of MBPT results is often influenced by error canceling. 

Going beyond the RPA for $W$ corresponds to the inclusion of vertex corrections beyond the GWA for $\Sigma_{xc}$, which has been an issue of intense research for decades. In agreement with results from literature\cite{Northrup1987,Northrup1989,Cazzaniga2012,Lischner2014}, here we find that passing from RPA to ALDA the QP bandwidth decreases by 0.22 eV, while the QP peaks increase slightly their width, implying a reduction of the QP lifetimes\cite{Cazzaniga2012} (see Fig. \ref{Fig:spf-RPA-TDLDA}).

\begin{figure}
\begin{center}
\includegraphics*[width=0.9\columnwidth, angle=0]{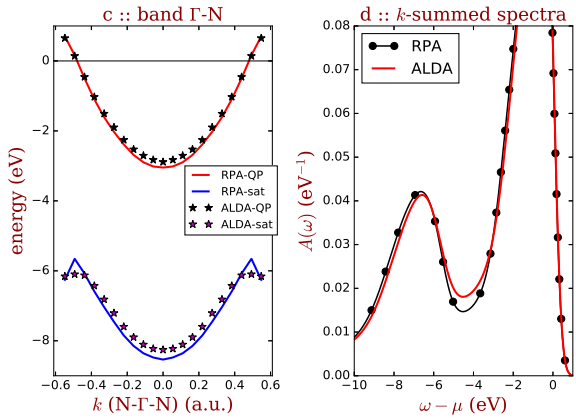}
\end{center}
\vspace{-15pt}
\caption{(a) Band and satellite dispersions and (b) $\vb{k}$-integrated spectral functions of sodium using RPA and ALDA screening.} 
\label{Fig:band-RPA-TDLDA}
\end{figure}

In line with the findings in the previous section, also in this case we find that the change in the screening affects more the satellites than the QPs [see Fig. \ref{Fig:band-RPA-TDLDA}(a)]: the \QP\ binding energy at $\Gamma$ is reduced by 0.22 eV due to the ALDA, while the satellite binding energy decreases by 0.37 eV. This leads to a decrease of the distance between the QP and the satellite of 0.15 eV, going from RPA to ALDA. 
Also in the $\vb{k}$-integrated spectral function [see Fig. \ref{Fig:band-RPA-TDLDA}(b)], both the increase of the QP width and a slight reduction of the binding energy of the center of mass of the satellite peak are apparent.
This means that using the ALDA instead of the RPA for the calculation of $W$, spectral functions are obtained in slightly better agreement with photoemission experiments. The comparison with experiment will be discussed more in detail in the next section.

\section{Comparison with  experiment}
\label{sec:experiment}

In Ref. \onlinecite{zhou-jcp2015} the spectral function of sodium valence was calculated using the TOC together with RPA screening without intraband contributions, the 5K lattice constant and a valence only pseudopotential. A discrepancy with experiment of almost one eV was found concerning the distance between the first plasmon satellite and the valence band, and it was speculated that the RPA might be at least partially responsible for this difference. As we have seen in the previous sections, the RPA does indeed lead to an overestimate of the QP-satellite distance of 0.15 eV, but several other effects add up: together with the effects of the lattice constant (0.15 eV), the core polarization (0.23 eV) the intraband contribution (0.2 eV) and the use of the RC instead of the TOC (0.11 eV), the total improvement amounts to the significant change of about 0.84 eV. 

While the comparison of peak positions with the experimental ones can be done on a quantitative level, the comparison of spectra including spectral weight and shapes is more delicate. For sodium, the ARPES data  of Jensen {\it et al.} \cite{Jensen1985,Lyo1988} 
displayed a bandwidth reduction due to interaction effects that was larger than predicted from HEG calculations, and a sharp peak at the Fermi energy for photon energies where there no  hole excitation should be possible in a single-particle picture. 
These experimental results gave rise to controversial interpretations \cite{Overhauser1987,Shung1987b}, with Overhauser\cite{Overhauser1985} proposing that the observed sharp peak close to the Fermi level was a signature of the existence of a charge-density wave, 
while Mahan and coworkers\cite{Shung1986,Shung1987,Mahan2000}
showed that  a careful description of the photoemission process itself was needed
to reconcile theory and experiment. 
This debate illustrates that for a detailed comparison with experimental photoemission spectra, the calculation of the intrinsic spectral function alone is not sufficient. However, the simulation of the photoemission process is a complex task itself. We therefore limit ourselves to a semi-quantitative comparison of spectra, following the here 
simplified approach used\footnote{The inclusion of the photoionisation cross sections following \cite{Guzzo-prl2011} was not possible here on the basis of available atomic data \protect\cite{TRZHASKOVSKAYA2001} as in Na atom the $3p$ shell is completely empty, contrary to the bulk case. } in Refs. \onlinecite{Guzzo-prl2011,zhou-jcp2015}.

\begin{figure}
\begin{center}
\includegraphics*[width=0.9\columnwidth, angle=0]{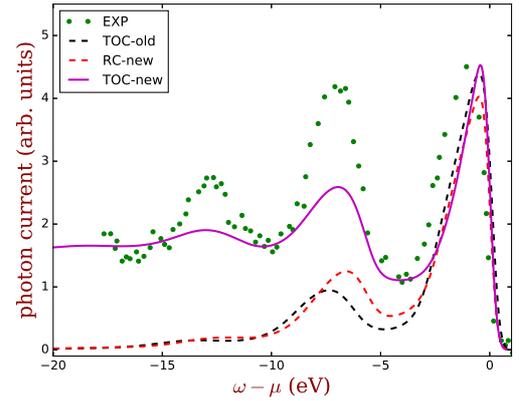}
\end{center}
\vspace{-15pt}
\caption{The black and red dashed curves are intrinsic TOC from Ref. \protect\onlinecite{zhou-jcp2015} (using the 5 K lattice constant, transitions from valence only, without intraband transitions, and RPA screening) and RC spectral functions (using room temperature lattice constant, including transitions from semi-core and intraband transitions, and ALDA screening), respectively. The black dashed curve has been shifted by 0.19 eV in order to align the QP with the red dashed curve.
The TOC (calculated as the RC) spectral function with extrinsic and interference effects, together with secondary electron background (magenta solid curve) is compared with experimental data from Ref. \protect\onlinecite{Hoechst1978} (green dots). } 
\label{Fig:spf-exp}
\end{figure}

The photoelectron leaving the sample undergoes scattering events: these extrinsic losses sum with the additional excitations induced by the photohole that are seen as satellites in the intrinsic spectral function. Moreover, the interaction of the photoelectron and the photohole produces an interference effect that partially cancels
with intrinsic and extrinsic contributions. In order to take into account these extrinsic and interference effects in the calculation of the photocurrent, we adopt the model of Hedin and coworkers \cite{Bardyszewski1985,Hedin1998}. Since this approach has been developed for the time-ordered formalism only, here we discuss these effects on the basis of the TOC spectral function. We also included the secondary electron background using a Shirley profile \cite{Shirley1972}, we multiplied the calculated spectral functions with a Fermi function for T = 300 K and  applied a Gaussian broadening of 0.255 eV corresponding to the experimental resolution\cite{Hoechst1978}. The final comparison between the calculated photocurrent for photon energy $h\nu=$ 1487 eV and the experimental data from Ref. \onlinecite{Hoechst1978} is shown in Fig. \ref{Fig:spf-exp}. 

The TOC intrinsic spectral function  (black dashed curve) is almost identical \footnote{The non-noticeable difference stems from the fact that in Ref. \onlinecite{zhou-jcp2015} the TOC was calculated based on a multi-pole sampling of $\Im\Sigma_{xc}$, while in this paper all the CEA results are produced using a new cumulant code based on a numerical integration of Eqs. \eqref{eq-toc-C} and \eqref{eq-rc-C}.\cite{computational-paper}}
to the results of Ref. \onlinecite{zhou-jcp2015}, with its overestimate of the QP-satellite distance of 0.8 eV, since it has been calculated using the same ingredients: RPA screening without intraband contributions, a valence only pseudopotential, and the 5K lattice constant. Our best intrinsic spectral function, namely the RC result obtained at the room temperature lattice constant and with ALDA screening including core polarization as well as intraband contributions, is given by the red dashed curve. 
The \QP\ peak of the two results is similar (the QP maximum of the red and black dashed curves is at $0.49$ and $0.68$ eV binding energy, respectively). However, it can be clearly seen that as outlined above, the binding energy of the first plasmon satellite in the new calculation is about 0.8 eV smaller than the old one, such reducing significantly the difference with respect to experiment. This can be better appreciated when the full photoemission experiment is simulated as explained above (magenta curve). Concerning the spectral shape and intensities, more work is needed: the experimental \QP\ is broader and slightly more symmetric than the calculated one, which may be due to several reasons, like the experimental resolution or temperature effects beyond the change in lattice constant.
Moreover also  the  photoionisation cross sections and the presence of the surface (with the corresponding surface plasmons) are known to play a role\cite{Hoechst1978}.
This also leads to an uncertainty in the relative normalization of the spectra, which are given in arbitrary units, and partially explain the apparent difference in the weight of the satellites. However, our method to simulate extrinsic and interference effects is admittedly quite crude, and one should not over-interpret the results.

\section{Conclusion}
\label{sec:conclusions}

We have presented a detailed study of the photoemission spectra of sodium and the \HEG , with a focus on plasmon satellites. This study is motivated by the increasing use of cumulant expansion approximations (CEAs) for the \textit{ab initio} calculation of photoemission spectra. While model studies in this context are numerous, many details concerning quantitative calculations remain to be elucidated.  

The main conclusion of the present work is the high sensitivity of satellites to many details of the calculations and, strictly related, to many details of the real material in experiments. Noticeable changes in the satellite positions occur due to thermal expansion and due to the effect of the crystal potential. Moreover, the semi-core polarization modifies the satellite positions. These effects are important to explain the measured spectra. \cite{Hoechst1978} On the computational side, improvements are also found by using TDDFT in the adiabatic local density approximation instead of the RPA for the calculation of screening. Moreover, the RC version of the CEA instead of the traditional TOC leads to a further small improvement of the satellite position, and creates electron removal satellites for spectral functions at $\vb{k} > \vb{k}_F$ which might be measurable if sufficient experimental resolution in $\vb{k}$ and energy is available. A fully quantitative comparison with experiment is beyond the scope of this work, since photoemission contains many effects that go beyond the intrinsic spectral function. 
In particular, the inclusion of extrinsic and interference effects has up to now only been done in a very approximate way, and with a prescription that is limited to the TOC.  However, our study yields detailed insight about interesting features of the intrinsic spectral functions and about the care that is needed in the calculations, and it highlights the potential impact of studies of the satellite part of photoemission spectra for the understanding of materials.

\acknowledgements
The research leading to these results has received funding from the European Research Council
under the European Union's Seventh Framework Programme (FP/2007-2013)
/ ERC grant agreement no. 320971 
and from a Marie Curie FP7 Integration Grant. 
Computation time was granted by GENCI (Project No. 544). 
JJR acknowledges hospitality by the Ecole Polytechnique,
with financial support by the Labex NanoSaclay and the chaire X-ESPCI-Saint Gobain ``Sciences des Mat\'eriaux et Surfaces Actives''. 
JJK and JJR are supported in part by the US DOE BES
Grant DE-FG02-97ER45623. We acknowledge fruitful discussions with Marco Cazzaniga.

\appendix
\section{Computational details \label{sec: computation-details}}
\label{sec:compdet}
We carry out energy self-consistent GW (EscGW) calculations (updating $G$, but keeping $W$ fixed) using a plane wave basis and norm-conserving Troullier-Martins-type pseudopotentials \cite{pseudop-PRB1993} as implemented in the ABINIT code \cite{Abinit-code}. The Brillouin zone (BZ) of sodium and \HEG\ are both sampled using a $20\times20\times 20$ grid mesh that yields 145 inequivalent k-points in the irreducible Brillouin zone (IBZ) for sodium, and 726 k-points for \HEG, since sodium is face-centered cubic and our \HEG\ is simulated using a simple cubic structure. A smearing temperature of 0.005 Ha was used for all the calculations. This is a fictitious temperature that only serves as a computational trick to speed up the k-point convergence, which explains why we can still use a standard time-ordered formalism in the GW calculation (besides the fit of the intraband contribution to the screening, see below).

The plane-wave cutoff of the LDA ground-state calculation was 6 Ha for the {\HEG}, 16 Ha for sodium with valence electrons only, and 200 Ha for sodium containing core electrons. The converged parameters for the calculation of screening and self-energy are reported in the table \ref{tab:table-converge}, where the first part contains the parameters for screening calculation and the second part is for the self-energy calculation. The Lorentzians in both $\chi_0$ and $\Sigma$ (e.g., $\eta$ in Eq. \eqref{Eq:model-sigma-Hedin}) are chosen to be 0.1 eV in all calculations. 

In Tab. \ref{tab:table-converge}, nband refers to the number of bands, npwwfn and npweps are the number of plane waves representing the wave functions and the dielectric matrix, respectively, and nfreqim, nfreqre are the number of imaginary and real frequencies, respectively. The maximum real frequency is represented by freqremax. The number of plane waves for the exchange part of the self-energy is named npwsigx.

The intraband transitions in the dielectric function for $\vb{q}=0$ are taken into account approximately using $\epsilon_{intra} = 1 - \omega_p^2/[\omega(\omega+\imi \eta)]$ \cite{Cazzaniga2012,Cazzaniga-PRB2010-intraband} where the parameters $\omega_p$ and $\eta$ are fitted on the calculated retarded loss function for small $\vb{q} \neq 0$.  
\begin{table}[h]
\caption{\label{tab:table-intraband}%
Parameters used in the intraband transitions
}
\begin{ruledtabular}
\begin{tabular}{ l c c}
\textrm{systems}&
\textrm{$\omega_p$ (eV)} &
\textrm{$\eta$ (eV)}\\
\colrule
\textrm{HEG} & 6.04& 0.1\\
\textrm{Na-5k (valence)} &5.95& 0.14\\
\textrm{Na-293k (valence)}& 5.83&0.136\\
\textrm{Na-293k-core-rpa} &5.55& 0.135\\
\textrm{Na-293k-core-alda}&5.48& 0.145\\
\end{tabular}
\end{ruledtabular}
\end{table}

The spectra of the cumulant expansion approximations are calculated using our cumulant code.\cite{computational-paper} The cumulant code takes the outputs of the GW calculation from the ABINIT code. In particular, we evaluate Eqs. \eqref{eq-toc-GF}, \eqref{eq-toc-C} for the calculation of the time-ordered cumulant, and Eq. \eqref{eq-rc-C} in the retarded cumulant calculation. 

\begin{table}[h]
\caption{\label{tab:table-converge}%
Parameters in the GW calculations
}
\begin{ruledtabular}
\begin{tabular}{ l c c c}
\textrm{Parameters}&
\textrm{HEG}
&\textrm{Na (valence)}
&\textrm{Na (core)}\\
\colrule
nband& 30&60& 60\\
npwwfn& 50&100&1500 \\
npweps &50&50&50\\
nfreqim& 25 &10 & 10\\
nfreqre&225 & 150& 150\\
freqremax&25 eV& 25 eV&25 eV \\
\colrule
nband&30 & 60 & 60 \\
npwwfn& 50& 200& 9000\\
npwsigx & 50& 200& 9000 \\
\end{tabular}
\end{ruledtabular}
\end{table}

\bibliographystyle{apsrev}
\bibliography{na_plasmon.bib}

\end{document}